\documentclass[]{article}
 
\usepackage[]{graphics}
\usepackage[]{psfig}

\def\ltsima{$\; \buildrel < \over \sim \;$}
\def\simlt{\lower.5ex\hbox{\ltsima}}
\def\gtsima{$\; \buildrel > \over \sim \;$}
\def\simgt{\lower.5ex\hbox{\gtsima}}                                           

\begin{document}
 
\begin{center}
{\bf \large  The CORALS Survey: A Review and Progress Report on The Search
for Dust Obscured Quasar Absorption Line Systems.}
\end{center}
\vspace{0.5cm}
 
\begin{center}
 
{\it Sara L. Ellison$^{\, 1,2}$, Max Pettini$^{\, 3}$, 
Chris W. Churchill$^{\, 4}$, 
Isobel M. Hook$^{\, 5}$, Sebastian Lopez$^{\, 6}$, 
Samantha A. Rix$^{\, 3}$, Peter Shaver$^{\, 1}$, Jasper V. Wall$^{\, 7}$,
Lin Yan$^{\, 8}$ 
}
 
\vspace{0.5cm}
\end{center}
 
\noindent $^{1}$European Southern Observatory\\      
\noindent $^{2}$P. Universidad Catolica de Chile, Santiago, Chile\\
\noindent $^{3}$Institute of Astronomy, Cambridge, UK\\
\noindent $^{4}$Penn. State University, State College, USA\\
\noindent $^{5}$Gemini Observatory, Oxford, UK\\
\noindent $^{6}$Universidad de Chile, Santiago, Chile\\
\noindent $^{7}$Dept. of Astrophysics, University of Oxford, UK\\
\noindent $^{8}$SIRTF Science Center, Pasadena, California, USA\\

\section{Ashes to Ashes, Dust to Dust}

The presence of dust represents a
perennial problem in many fields of astrophysics, ranging from
its role in the supernova-determined distance scale to
its depletion of refractory elements in the interstellar medium (ISM).
However, dust is very much a necessary evil since it regulates the 
temperature of the ISM and
provides a shield against UV radiation, as well as nucleation sites for
the formation of H$_2$.   Despite
its ubiquitous astrophysical impacts, the formation of dust, and even
its composition, remain poorly understood.  Although it is thought that
the majority of dust is formed in the outer regions of cool AGB
stars, it is also possible that supernovae and AGN may contribute
significantly to the dust budget at high redshifts.
Given the widespread evidence
for significant amounts of dust even at early times, 
it is hard to escape the possible consequences
of depletion and extinction effects on astronomical observations.
 
The study of QSO absorption line systems is a field in which dust
continually plagues our interpretation of the data.  This technique
uses relatively bright, yet distant, quasars as background sources to study
intervening gas clouds, which imprint their signatures on 
the quasar spectrum allowing detailed study of their composition and
structure.  Echelle
spectrographs such as UVES on the VLT are now, almost routinely,
providing exquisite data that permit accurate measurements of gas
phase abundances in galaxies and the intergalactic medium 
out to very high redshifts.  However, the interpretation of these
measurements and their application to galaxy evolution scenarios
is hindered by depletion from the gas phase of refractory elements,
effectively `hiding' some of the chemical elements from view.
Although techniques are being developed to circumvent the effect
of depletion, 
this is a challenging prospect when
little is known about dust composition outside the Local Group.

A perhaps more fundamental concern is that the surveys that
search for absorption line systems may themselves be incomplete due to 
dust in intervening absorbers.  If the internal extinction of
absorption systems is sufficiently large, then quasars behind such systems may
be missed by magnitude limited optical surveys.
Junkkarinen et al (2003) have recently underlined this concern with
the detection of strong interstellar dust features in a $z \sim 0.5$
damped Lyman $\alpha$ system.  Indeed, Fall \& Pei (1993) have 
used models of dust obscuration to
estimate that between 30 and 70\% of QSOs could be missed in present
surveys due to this very effect.

\section{Towards a Complete View of Absorption Line Galaxies}

The Complete Optical Radio Absorption Line System (CORALS) survey
was designed to provide a quantitative answer to concerns about
absorption line survey dust bias.  The aim, simply put, was to compile 
a sample of QSOs selected at radio wavelengths with 
\textit{no optical magnitude limit} from which absorption line statistics
could be determined.  The parent sample for this survey is
the Parkes quarter-Jansky (PQJ) sample (Jackson et al. 2002) which
contains 878 flat spectrum radio sources observed at 2.7 and 5.0
GHz.  An important feature
of the PQJ sample is the extensive follow-up imaging campaigns that
have resulted in optical identifications and classifications
for all but 9 of the sources.  Selected at
wavelengths that are immune to dust extinction, but with essentially
complete optical identifications\footnote{The 9 unidentified sources
were due to mis-identifications in earlier samples, not due to
excessively faint optical magnitudes.  Moreover, only one of these
falls within our declination range and above its survey flux-density 
limit.  The unidentified objects in Jackson et al (2002) therefore have
negligible effect on the results of the CORALS survey.}, the PQJ sample
provides us with an excellent opportunity to address the possibility
of dust bias in previous magnitude limited absorption system surveys.
Although a large spectroscopic campaign was undertaken for many of
PQJ sample, these were
obtained at low resolution for the purpose of object classification
and redshift determination and are not suitable for absorption
system studies.  Therefore, over the last 5 years, we have been
pursuing an active observing campaign that has so far logged some 30
nights on telescopes over 4 continents to address issues associated 
with obscuration bias. 
 
\section{The First CORALS Survey }

The initial goal of the CORALS survey was to assess the possible bias
in samples of high redshift damped Lyman $\alpha$ systems (DLAs),
the highest column density absorbers associated with galaxy scale
systems.  The sample for this survey consisted of 
the 66 $z_{\rm em} > 2.2$ QSOs from the PQJ survey which had magnitudes 
as faint as $B=23.5$; complete results from this survey have been published
in Ellison et al (2001). Moderate
resolution spectra were obtained for this sample using a combination
of the AAT and ESO facilities, with the VLT playing the vital role
of observing the faintest targets.  Although the aim of CORALS~I was to
provide a complete survey, the faintest targets were the linchpin of
the project, since these are the QSOs which will have been previously 
excluded by magnitude limited optical samples.  The total path length
covered by CORALS~I is $\Delta z \sim 55$, over which a total of 22
DLAs were identified.  

\begin{figure}
\resizebox{14.0cm}{!}{\includegraphics{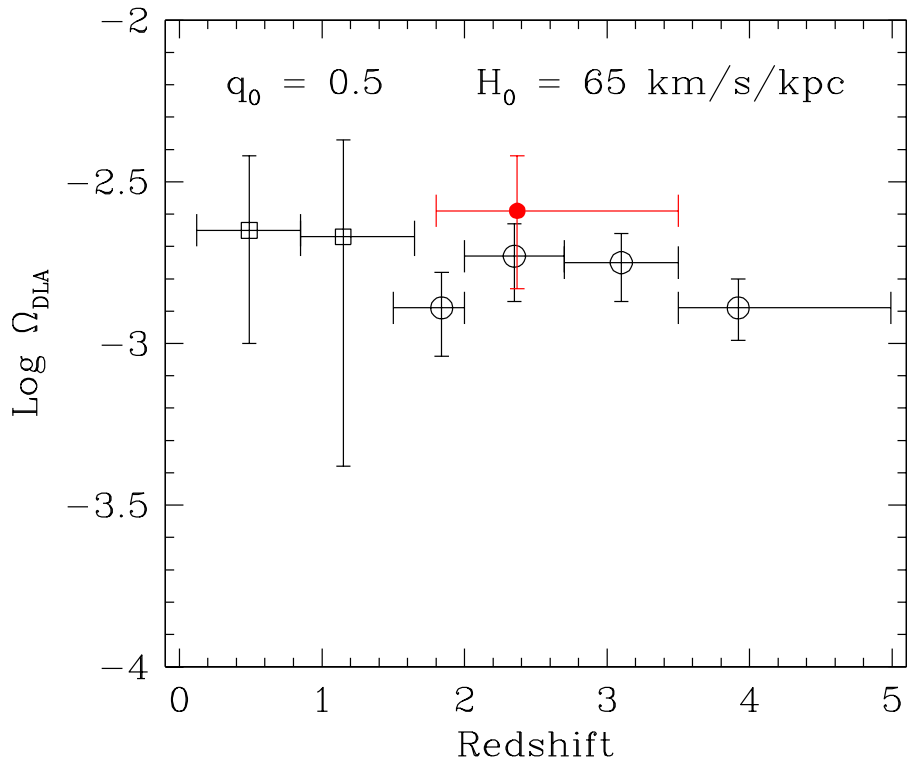}}
\caption{The mass density of neutral gas, $\Omega_{\rm
DLA}$, in DLAs.  Open circles and squares are measurements
from the latest compilations by P\'{e}roux et al. (2001) and
Rao \& Turnshek (2000) respectively.  The solid red circle is
the value from the CORALS~I survey presented here for the
redshift interval $1.8 < z_{abs} < 3.5$.  These results show that
for $z > 2$ the effect of dust bias has caused the under-estimate
of $\Omega_{DLA}$ by at most a factor of two.}
\label{omega}
\end{figure}

\begin{figure}
\resizebox{14.0cm}{!}{\includegraphics{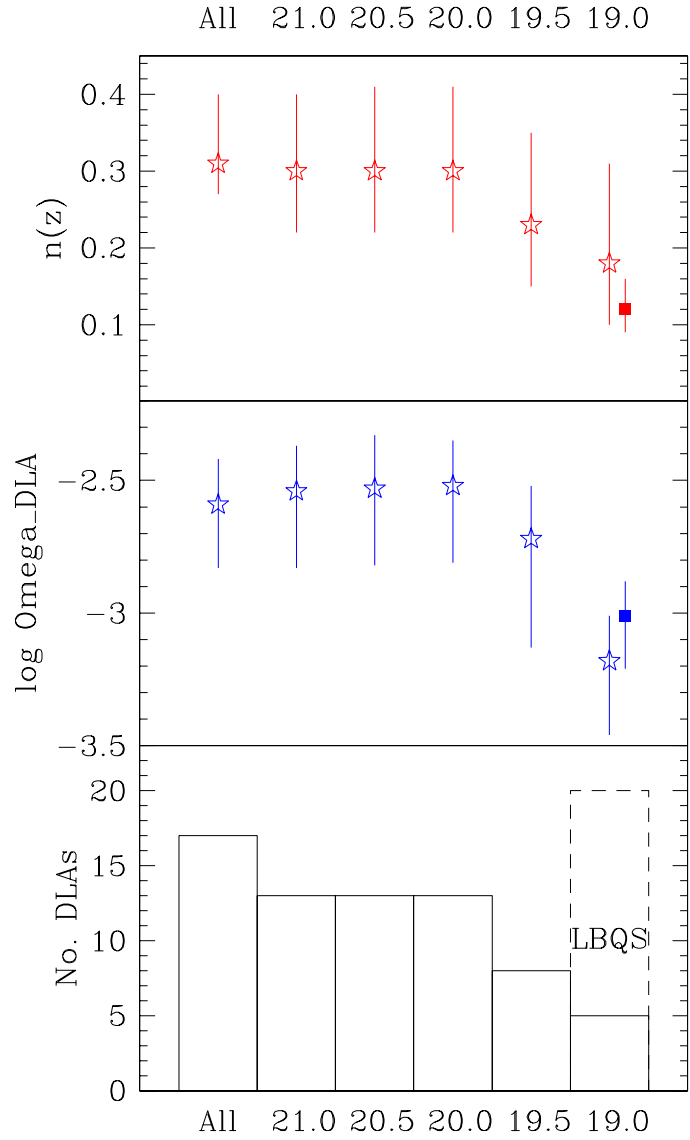}}
\caption{Cumulative DLA statistics for CORALS~I as a function of $B$
band magnitude show
a possible trend with QSO magnitude with a lack of DLAs in bright 
optically-selected samples like the LBQS.  Open stars represent
values from CORALS~I, whereas the solid squares include absorbers
from the LBQS (included to reduce the error bars of the bright
QSO bin).}
\label{nhisto}
\end{figure}

In order to compare CORALS~I with previous magnitude limited surveys,
we assess two main statistics.  The first, $n(z)$ is the number density; 
this is a simple tally of the number of absorption systems per 
unit redshift.    
The second statistic, $\Omega_{DLA}$, is a measure of the 
total neutral gas content of the DLA population, expressed as a
fraction of the closure density of the universe (for a more detailed
explanation of these quantities, see Ellison et al. 2001).  We determine
$n(z)$=0.31 at a mean redshift of $\langle z \rangle = 2.37$ in CORALS~I, 
in good agreement (within 1$\sigma$) with previous surveys. 
Similar reasonable agreement
is determined for $\Omega_{DLA}$ (see Figure \ref{omega}), although the
error bars permit an underestimate of up to a factor of $\sim$2 by
previous surveys, which typically include QSOs down to magnitudes 
of $V \sim$ 19--20.  However, there is evidence that statistics may
depend on survey magnitude for surveys that are only complete to significantly
brighter magnitudes.  In Figure \ref{nhisto} we plot the cumulative
statistics of the CORALS~I survey as a function of $B$ band magnitude,
and include the Large Bright Quasar Survey (LBQS) to improve
the statistics of the brightest QSOs where CORALS has poor coverage.  
As first pointed out by
Ellison et al. (2000) in a previous Messenger article, fewer
DLAs are found towards brighter QSOs (e.g. the LBQS, $B<19$) than fainter
subsets (e.g. CORALS $B>20$), and the total gas content is
also somewhat lower, although the
error bars remain large.  Such a trend is supported by the DLA
survey conducted using the Hamburg-ESO (HE) sample of bright QSOs, in
which $\Omega_{DLA}$ is an order of magnitude lower than for
CORALS (Smette et al., in preparation).  The precise dependence
of DLA statistics on survey magnitude limit not only has an important 
application in the design of future surveys, but also has implications
for the large datasets being reaped from surveys such as 2dF
and SDSS.  These surveys are
sufficiently large (with $\Delta z$ reaching several thousand) that 
error bars will be much less dominated by redshift coverage, so that 
observational biases, even subtle ones, will be important.

\section{CORALS~II:  Extension to Lower Redshift} 

The preliminary results from CORALS~I indicate that at $2 < z < 3$,
dust does not seem to play a significant role in `hiding' DLAs from
previous surveys, at least when QSOs with magnitudes $V\sim20$
can be reached.  However, it might be expected that biasing becomes
more severe towards lower redshifts, since the bulk of star formation has
already taken place by $z \sim 1$ (Steidel et al. 1999).  
With most of the star formation completed, we may expect the ISM of 
galaxies to exhibit pronounced chemical (and therefore, plausibly, dust) 
evolution at low $z$.
Dust obscuration could thus be invoked to posit a population of 
`missing', dust obscured DLAs, at low redshift, 
which could explain the lack of metallicity
evolution at $z < 1$ seen by Pettini et al. (1999).

Observationally, it is much more challenging to extend CORALS to $z<1.5$,
due to the onset of the atmospheric cut-off which renders detection
of low redshift Ly$\alpha$ impossible from the ground.  Although
large DLA surveys have been conducted with HST 
and other space telescopes, these
are very expensive in terms of telescope resources.  Moreover, current
HST instrumentation restricts surveys to bright magnitudes, and we
have seen that absorption statistics may depend to some extent on
magnitude cut-off.  Therefore, we have designed CORALS~II to select
absorption galaxies via Mg~II and Fe~II lines -- strong metal lines
associated with galaxy halos that have transitions observable in the optical
regime down to $z \sim 0.3$ (Bergeron \& Boisse 1991).  By selecting
systems with strong Mg~II and Fe~II absorption, we can
efficiently pre-select likely DLAs (Rao \& Turnshek 2000).  

\begin{figure}
\centerline{\rotatebox{0}{\resizebox{0.9\textwidth}{!}
{\includegraphics{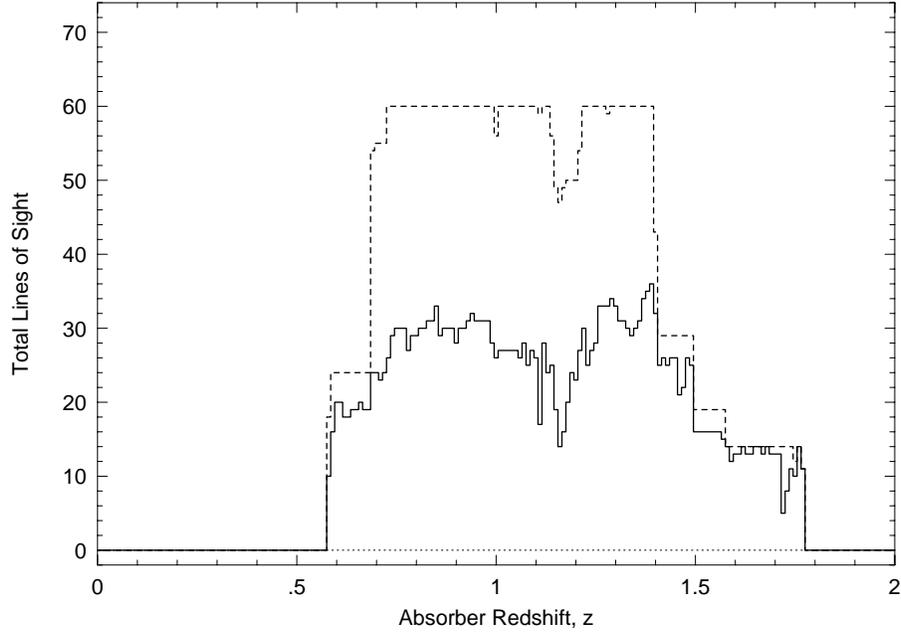}}}}
\caption{Total number of quasars in the CORALS~II survey which reach a 
given rest frame equivalent width
detection limit as a function of redshift.  Solid line is 0.3 \AA
and the dashed line is 0.6 \AA.  For
a total of 60 QSOs observed so far, we achieve a limit of 0.6 \AA\
for essentially all QSOs between $0.7 < z < 1.4$ (the small
dip at $z\sim1.2$ is due to incomplete wavelength coverage in
some of the spectra).}
\label{gz}
\end{figure}

\begin{figure}
\centerline{\rotatebox{0.0}{\resizebox{0.9\textwidth}{!}
{\includegraphics{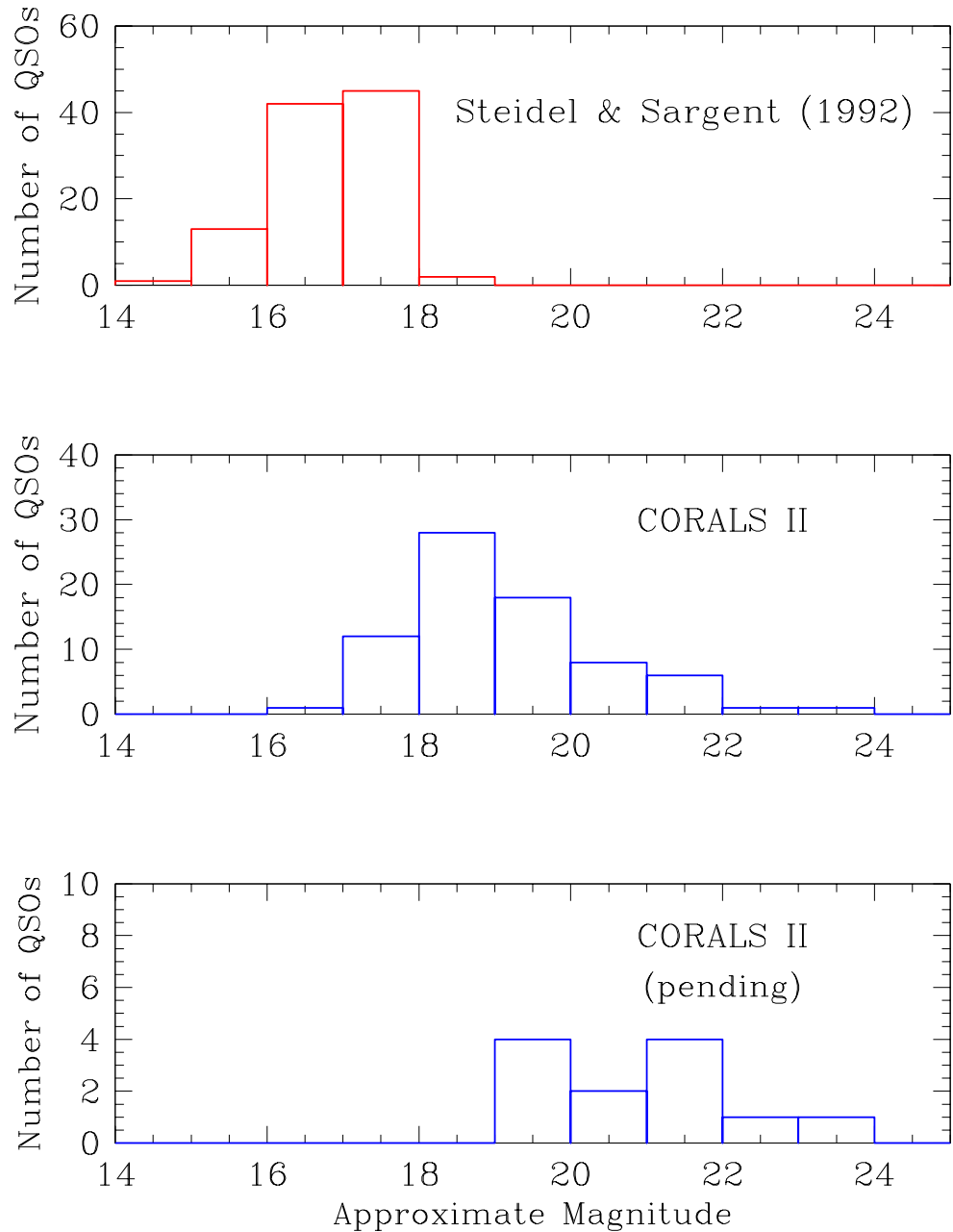}}}}
\caption{Comparison of the QSO magnitudes for the Steidel \& Sargent
(1992) Mg~II survey and CORALS~II.  The bottom panel shows the final
targets that are still pending observation.  The SS92 survey is
effectively a `bright' QSO sample, whereas CORALS~II is optically
complete and includes QSOs up to 250 times fainter than the SS92
limit. }
\label{SS92}
\end{figure}

CORALS~II, a complete survey for Mg~II absorbers with $0.5 < z < 1.5$
is currently nearing completion; out of 75 QSOs, we have so far observed
some 60 targets, the rest pending observation (mostly
with FORS on the VLT) in Period 71.  The QSO sample is again based on 
the PQJ flat spectrum quasar sample, although we have now preferentially 
selected $z_{\rm em}<2.5$ targets so that Mg~II will fall redwards of 
the Ly$\alpha$ forest.  In the majority of cases, we also cover 
Fe~II $\lambda$2600 and usually also Mg~I $\lambda$2853.
Our aim is to be complete
down to an observed 3$\sigma$ equivalent width threshold of 0.5 \AA\ for 
Mg~II, although in most cases we achieve limits significantly beyond this.
Figure \ref{gz} shows the number of QSOs in which we can achieve various 
sensitivity limits as a function of redshift, based on the data obtained
so far. Up to this point, we
have a redshift path coverage $\Delta z \sim 50$ for an equivalent
width limit of 0.5 \AA, which will increase to approximately
60 by the end of the survey.  We have so far detected 28 Mg~II
absorbers with $EW(Mg~II \lambda 2796) \ge 0.5$ \AA\ and a further
10 with $EW(Mg~II \lambda 2796) \ge 0.3$ \AA.   We can compare these
statistics with the landmark survey of Steidel \& Sargent (1992, hereafter 
SS92) performed with the Palomar 5-m telescope on a sample of QSOs with
$15 < V < 18$.  We determine a number density of absorbers that is,
considering the error bars, marginally lower than SS92; for an equivalant 
width threshold of $EW>$0.6 \AA\ (the limit used by SS92) we determine
$n(z)=0.46\pm0.10$ (at $\langle z \rangle = 1.08$) compared with $0.65\pm0.07$ 
at a similar mean redshift for SS92.   \textit{This is the opposite to what
we would expect if a dust bias is at work.} In Figure \ref{SS92}
we show the distribution of optical magnitudes for the SS92 survey 
compared with CORALS~II as it currently stands,
as well as the complete sample which is still pending completion.
Although these magnitudes have error bars which may exceed 0.3 mags
(and the CORALS radio-loud QSOs are expected to be highly variable),
the basic picture is that the Steidel \& Sargent (1992) sample occupy a locus
of brighter magnitudes than CORALS.  In fact, the SS92 magnitude range
is similar to that of the HE QSO survey.  
Whereas there seems to be a significant deficit of high redshift
DLAs in the HE bright QSO sample compared with our complete sample 
(Smette et al., in preparation),
we find tentative evidence of an excess of absorbers towards bright
QSO samples at intermediate redshift.  
This is suggestive of a lensing bias, whereby intrinsically
fainter QSOs are boosted by intervening galaxies and are included 
in brighter flux limited samples (e.g. Smette et al 1997). 
If we split the sample in half by emission redshift, the number density
for $z_{\rm em}>2.1$ is $n(z)=0.52\pm0.17$ and $0.41\pm0.13$ for lower
redshifts (for $\langle z \rangle \sim 1.1$ in both cases).  
Although these values are consistent within the large error bars,
the marginally higher $n(z)$ towards higher redshift QSOs is again
suggestive of lensing.  This is because the lensing efficiency 
(by intermediate redshift galaxies) is
higher for more distant QSOs (e.g. Bartelmann \& Loeb 1996).  
Larger samples, such as the SDSS and 2dF
surveys will be able to confirm this trend of $n(z)$ versus emission
redshift, even though they are confined to brighter samples.
We note that this is probably not an issue for
high redshift ($z_{\rm abs} > 2$) DLA surveys because of the low
lensing probability in this configuration.
Confirming the N(HI) of our complete Mg~II
sample, and thereby determining $\Omega_{DLA}$, will be an important 
test of whether a bright magnitude cut-off induces a bias in the
determination of the neutral gas density in DLAs at low $z$.
Such a bias is predicted to overestimate $\Omega_{DLA}$ (Smette et al.
1997) because the line of sight preferentially passes through
the inner part of the lensing galaxy.

\section{Along the Way...}

Sizeable surveys of any kind often produce spin-off projects which either
focus on a few unusual objects, or can exploit large datasets
to study the properties of subsets of the data.  We briefly review two
such spin-offs from the CORALS survey.

Traditional DLA surveys have excluded DLAs within $\sim$3000 km/s of
the QSO due to proximity effects and the
possibility that the absorber may be associated with the QSO itself.
However, M\o ller, Warren \& Fynbo (1998) have argued that, at least 
in some cases, proximate
DLAs (PDLAs) are likely to be the same beast as intervening
absorbers, based on their typical metallicities and lack of
high ionization lines.  If correct, we can use PDLAs as a probe of
galaxies that are clustered around QSOs at high redshift.
By comparing the $n(z)$ in the radio-loud quasar CORALS sample,
Ellison et al. (2002) found 4 times the number of PDLAs in CORALS~I
than towards the radio-quiet sample of Peroux et al (2001).
Although this result is only significant at the 2$\sigma$
level, it supports the suggestion that galaxies cluster
preferentially near radio-loud QSOs.

\begin{figure}
\centerline{\rotatebox{0.0}{\resizebox{0.9\textwidth}{!}
{\includegraphics{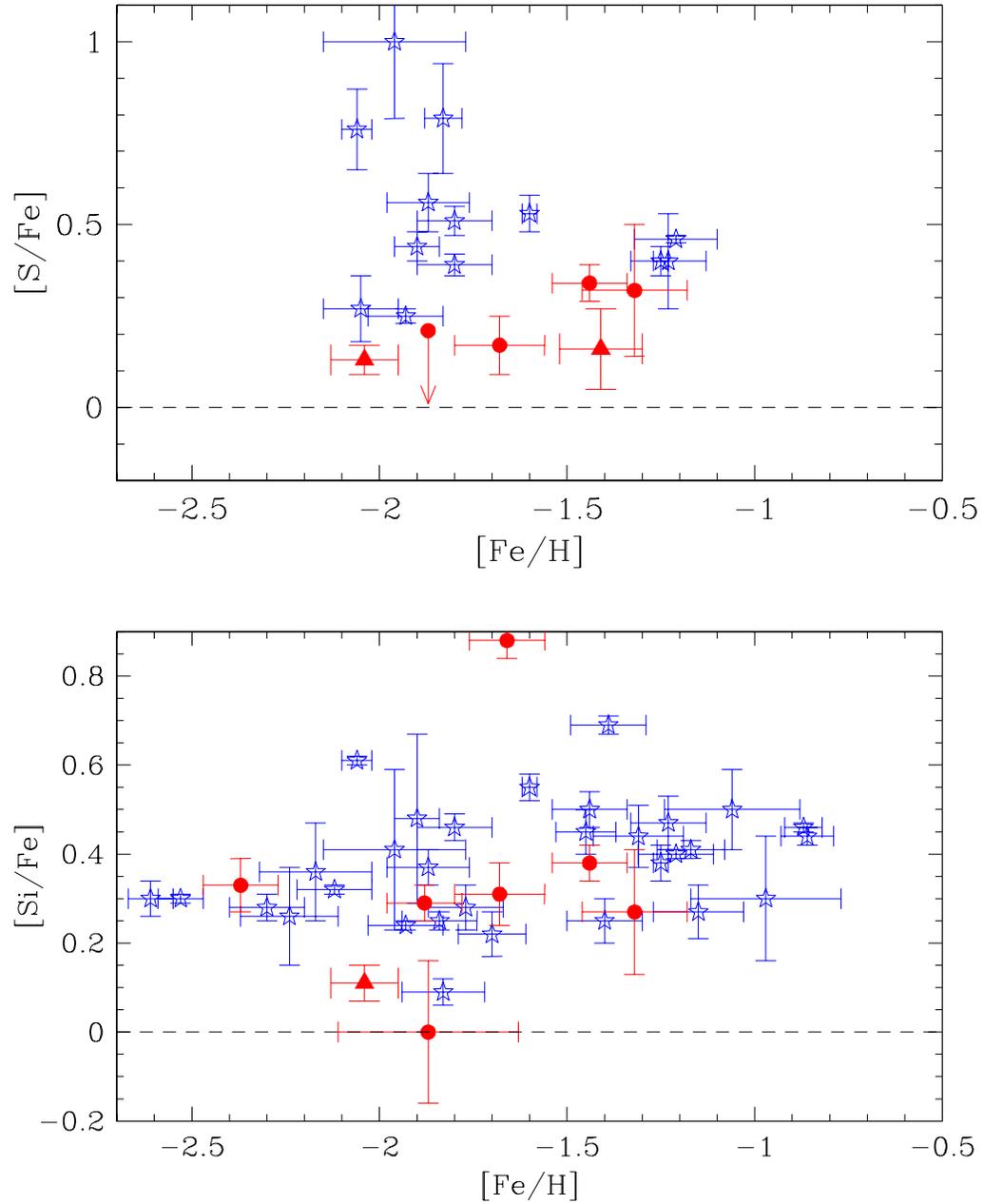}}}}
\caption{$\alpha$/Fe ratios for MDLAs (solid red circles), DLAs in fields
with known galaxy neighbours (solid red triangles) and single DLAs taken
from the literature (open blue stars).  DLAs with nearby galaxies both
in the field, and seen in absorption (MDLAs) have systematically
lower $\alpha$/Fe, a trend particularly obvious in the [S/Fe] ratio. 
See Lopez \& Ellison (2003) for further discussion.}
\label{alpha}
\end{figure}

A second spin-off to have been born of CORALS is the study of multiple
DLAs (MDLAs).  Lopez \& Ellison (2003) define an MDLA as two or
more absorbers with log N(HI) $>$ 20.0 with velocity separations
$500 < \Delta v < 10000$ km/s.  One of the DLAs discovered during
the CORALS~I campaign, Q2314$-$409, conforms to this definition and 
was the first to be studied at high resolution (Ellison \& Lopez 2001).    
The abundances determined from a UVES spectrum show a propensity towards
low $\alpha$/Fe (where $\alpha$ elements include such metals as Ca,
Si, S and O) for MDLAs compared with single absorbers, a
result more recently backed up by Lopez \& Ellison (2003), see
Figure \ref{alpha}.  Having ruled out systematic effects such
as ionisation or atypically low dust depletion, we have suggested 
that this abundance pattern could be due to low star formation
efficiencies, possibly linked with environment (assuming that MDLAs
are not just chance alignments, as indicated by the low statistical
probability of such an event).  To confirm this
hypothesis will require a larger abundance study of MDLAs, as well
as imaging campaigns to determine whether galaxy excesses exist
in these fields.

\section{Future Work}

Compiling the spectroscopic samples to search for absorption systems
has represented a significant investment of telescope time over
the last 5 years and has yielded the first quantitative estimate of
dust bias in magnitude limited surveys.  Building on this investment, 
several follow-up projects are already underway that capitalise
on the groundwork we have so far completed.  One of the most important
of these is to conduct an unbiased census of metallicity amongst
the high redshift DLA sample.  Although $\Omega_{DLA}$ appears
not to have been seriously biased by selection effects, we currently
have no measurement of the metal content of the absorber sample.  Using
UVES on the VLT, and with supplementary observations with MIKE on
Magellan and ESI on Keck, we have almost completed the observations
that will eventually yield the first unbiased metallicity measurement of
DLAs at high redshift.  Results from this will be published next year.

We have also initiated a program to study the optical-IR colours
of CORALS QSOs, using a combination of CTIO and ESO facilities.
Our aim is to obtain almost simultaneous optical and IR photometry
(a high level of simultaneity is required due to the rapid
variation associated with radio loud QSOs) in order to investigate
whether a significant amount of reddening of the background source
can be induced by intervening absorption galaxies.  A small
effect on the optical colours of the 2dF sample has been seen
by Outram et al (2001), but adding an IR band will greatly
increase the baseline over which reddening can be detected.

Further in the future will come HST confirmation of the H~I column
density of Mg~II selected absorbers from CORALS~II.  Although
conclusions on survey completeness can already be drawn from
the ground-based spectroscopic campaign described here, measuring
the N(H~I) is essential for determining whether or not these
are \textit{bona fide} DLAs.  Moreover, it is only with these
measurements that we can determine $\Omega_{DLA}$ in a complete
sample, and compare the gas content with the brighter QSO samples
that currently dominate the low redshift measurements (e.g.
Rao \& Turnshek 2000, see Figure \ref{omega}).  The efficiency
of STIS is such that obtaining spectra for all our Mg~II
absorbers is not feasible (all our ground-based spectroscopy has
been done with at least 4-m telescopes).  However, the installation of
the Cosmic Origins Spectrograph (COS) on HST, currently scheduled
for the start of 2005, will provide a facility capable of obtaining
moderate resolution spectra sufficient to determine N(H~I) for
the bulk of our sample.

\section{Acknowledgements}

We are extremely grateful to the continued support of the various time
allocation committees who have granted time for CORALS related projects
over the last 5 years and to the various observatory staff who have
facilitated work on site.

\begin{footnotesize}
\vspace{1cm}

\noindent {\it \large References}

\vspace{0.5cm}

\noindent Bartelmann, M., \& Loeb, A., 1996, ApJ, {\bf 457}, 529\\
\noindent Bergeron, J., \& Boisse, P., 1991, A\&A, {\bf 243}, 344\\
\noindent Ellison, S. L., \& Lopez, S., 2001, A\&A, {\bf 380}, 117\\
\noindent Ellison, S. L., Yan, L., Hook, I., Pettini, M., Shaver, P.,
Wall, J., 2000, ESO Messenger, {\bf 102}, 23\\
\noindent Ellison, S. L., Yan, L., Hook, I., Pettini, M., Wall, J., Shaver, P.,
2001, A\&A, {\bf 379}, 393\\
\noindent Ellison, S. L., Yan, L., Hook, I., Pettini, M., Wall, J., Shaver, P.,
2002, A\&A, {\bf 383}, 91\\
\noindent Fall, S. M., Pei, Y. C., 1993, ApJ, {\bf 402}, 479\\
\noindent Jackson, C. A., Wall, J. V., Shaver, P. A., Kellermann, K. I.,
Hook, I. M., Hawkins, M. R. S.,et al., 2002,  A\&A, {\bf 386}, 97\\
\noindent Junkkarinen et al. 2003, ApJ, submitted\\
\noindent Lopez, S., \& Ellison, S. , 2003, A\&A, accepted, astro-ph/0303441\\
\noindent M\o ller, P., Warren, S. J., Fynbo, J. U., 1998, A\&A {\bf 330}, 19\\
\noindent Outram, P. J., Smith, R. J., Shanks, T., Boyle, B. J., Croom, 
S. M., Loaring, N. S., Miller, L, 2001, MNRAS, {\bf 328}, 805\\
\noindent Pettini, M., Ellison, S., Steidel, C., Bowen, D., 1999, 
ApJ, {\bf 510}, 576\\
\noindent Peroux, C., Storrie-Lombardi, L. J., McMahon, R. G., Irwin, M.,
 Hook, I. M., 2001, AJ, {\bf 121}, 1799\\
\noindent Rao, S., Turnshek, D., 2000, ApJS, {\bf 130}, 1\\
\noindent Smette, A., Claeskens, J.-F., Surdej, J., 1997, New Astronomy
{\bf 2}, 53\\
\noindent Steidel, C.C., Adelberger, K.L., Giavalisco, M., Dickinson, M., \&
 Pettini, M. 1999, ApJ, {\bf 519}, 1\\
\noindent Steidel, C. C., \& Sargent, W. L, W., 1992, ApJS, {\bf 80}, 1\\
\end{footnotesize}

\pagebreak                                              
   
                     
\pagebreak                                              
                        
\end{document}